\documentclass[preprint,amsmath,amssymb,aps,showkeys,showpacs]{revtex4}
\usepackage[colorlinks=true, pdfstartview=FitV, linkcolor=blue, citecolor=red, urlcolor=magenta]{hyperref}
\usepackage{graphicx}
\usepackage{latexsym}
\usepackage{amsmath}
\usepackage{amsfonts}
\usepackage{amssymb}
\usepackage{verbatim}
\usepackage{dcolumn}
\usepackage{amssymb}
\usepackage{bm}
\usepackage{float}
\usepackage{subfigure}
\usepackage[english]{babel}
\newcommand{\be}{\begin{equation}}
\newcommand{\ee}{\end{equation}}
\newcommand{\bea}{\begin{eqnarray}}
\newcommand{\eea}{\end{eqnarray}}

\begin{document}

\newcommand{\NITK}{
\affiliation{Department of Physics, National Institute of Technology Karnataka, Surathkal  575 025, India}
}

\newcommand{\IIT}{\affiliation{
Department of Physics, Indian Institute of Technology, Ropar, Rupnagar, Punjab 140 001, India
}}

\title{Joule-Thomson Expansion of Regular Bardeen AdS Black Hole Surrounded by Static Anisotropic Quintessence Field}
\author{Rajani K.V}
\email{rajanikv10@gmail.com}
\NITK
\author{Ahmed Rizwan C.L.}
\email{ahmedrizwancl@gmail.com}
\NITK
\author{Naveena Kumara A.}
\email{naviphysics@gmail.com}
\NITK
\author{Md Sabir Ali.}
\email{alimd.sabir3@gmail.com}
\IIT
\author{Deepak Vaid.}
\email{dvaid79@gmail.com}
\NITK

%

\begin{abstract}
In the present paper, we investigate the required anisotropy of an exact regular Bardeen  black hole characterized by its mass $M$, the nonlinear parameter $g$, the quintessence field parameter $a$ in anti-de sitter spacetime with a static quintessence matter field. We also show that the relative pressure anisotropy, equation of state and the pressure depends on  radial coordinate, reflecting the required anisotropy for Bardeen black hole in the quintessence background. 

Next, we analyze the Joule-Thompson ($JT$) expansion of the black hole spacetime. Treating the cosmological constant as thermodynamic pressure $P$ and its conjugate quantity as thermodynamic volume $V$ we derive the equation of state connecting Hawking temperature and various black hole parameters. We study the $JT$ expansion in the regular Bardeen AdS black holes in the quintessence background through the analysis of inversion temperature and isenthalpic curves. We derive the $JT$ coefficient $\mu$, and use them to plot the inversion and isenthalpic curves. We discuss the effect of quintessence parameter $a$ and $\omega_q$ on the $JT$ coefficient and inversion temperature, especially with the case of $\omega_q=-1$ and $\omega_q=-\frac{1}{3}$. Our analysis shows that quintessence dark energy affects the inversion point $(T_i,P_i)$ .
\end{abstract}
\maketitle


\section{Introduction}
During the past three decades, the black hole phase transition in anti-de Sitter (AdS) space got wide attention since the proposal Hawking-Page phase transition \citep{Hawking:1982dh}. More recently, identifying the negative cosmological constant  as the pressure term and its conjugate quantity with the volume gave us more physical insights into the thermodynamic properties of black hole \citep{Kastor:2009wy}. In this extended approach, the close resemblance between the black hole phase transition and the van der Waal liquid-gas phase transition became apparent \citep{Chamblin:1999hg,Kubiznak:2012wp}. Interestingly, the phase structure exhibited in the black holes is quite meaningful in the view of AdS/CFT correspondence and in the context of the emergent gravity paradigm. Realizing the connection to the quantum gravity theories in these phenomena, there has been rapid progress in probing the van der Waals fluid-like behavior in various black holes. An important feature of the black hole phase transition is that all four critical exponents fall within the universality class. These developments lead to the new avenue in theoretical physics called the black hole chemistry. 

With these insights, most of the classical thermodynamic features of the real gas can be attributed to AdS black holes. Concept of heat engine, Joule-Thomson expansion, Clausius-Clapeyron relation, and Maxwell's equal-area law are constructed for asymptotically AdS black holes \citep{Johnson:2014yja,Okcu:2016tgt,Spallucci:2013osa}. In this article we focus on the Joule-Thomson expansion in charged AdS black holes, which is proposed by Okcu and  Aydner \citep{Okcu:2016tgt}. Joule-Thomson effect or throttling process happens in gas when it is allowed to move from the high-pressure region to the low-pressure region without any change in enthalpy. In the extended approach to thermodynamics the black hole mass is understood as the enthalpy. During the Joule Thomson expansion the  mass remains constant, so it is also called as isenthalpic process. This process will result in heating or cooling in the final phase. The JT coefficient $\mu$, the gradient of temperature with pressure $(\mu=\frac{\partial T}{\partial P})$ determines the temperature change in the final phase. The conditions, $\mu>0$ implies a cooling and $\mu<0$ heating phases. These two curves, isenthalpic  (constant enthalpy) and inversion curves are used to study the JT expansion of the gas. The point at which $\mu=0$, denotes the position of inversion points $\left(T_i,P_i\right)$. The inversion points distinguish heating and the cooling phases and the inversion curve is the locus of inversion points for different isenthalpic curves. 
Following the work of Okcu and Aydner, there were many studies on JT expansion in various black holes \citep{Okcu:2017qgo,Ghaffarnejad:2018exz,Mo:2018rgq,Chabab:2018zix,Mo:2018qkt,Zhao:2018kpz,Lan:2018nnp,AhmedRizwan:2019yxk,Ghaffarnejad:2018tpr,Kuang:2018goo,Cisterna:2018jqg,Haldar:2018cks,Li:2019jcd,Guo:2019gkr,Yekta:2019wmt,Nam:2019zyk,Rostami:2019ivr,Lan:2019kak,Ranjbari:2019ktp,Guo:2019pzq,Sadeghi:2020bon}.  

The weak cosmic censorship hypothesis insists that the presence of singularity must be hidden from the far observer by a membrane called the event horizon \citep{Hawking:1969sw, Hawking:1973uf}.  All the spacetime solutions for Einstein's equations respect this conjecture. However, there are black hole spacetimes without singularity at the origin and possessing an event horizon. In regular black holes, the singularity at the origin is removed by a repulsive de-Sitter core. For the first time , inspired by the works of  Sakharov \citep{1966JETP...22..241S} and Gliner \citep{1966JETP...22..378G}, Bardeen came up with a regular black hole solution \citep{bardeen1968non}. Following the idea of Bardeen, later several regular solutions for Einstein's equations coupled to nonlinear electrodynamic sources were found \citep{Hayward:2005gi, AyonBeato:1998ub, AyonBeato:2000zs}. The  black hole thermodynamics of regular black holes are studied in detail by several authors \citep{Man:2013hpa, Man:2013hza,Aros:2019quj,Nam:2018ltb,PhysRevD.98.084025,Kumar:2018vsm}.

In recent times, the black holes with quintessence have attracted wide attention due to the importance of quintessence in cosmology. The quintessence is one of the most important candidate to explain the accelerated expansion of the universe. The observational cosmological data supports  the equation of state $p_q=\omega_q \rho _q$,  where the quintessence state parameter varies as $-1< \omega_q < -1/3$. The energy density  for quintessence has the form $\rho _q=-\frac{a}{2}\frac{3\omega_q}{r^{3(\omega_q+1)}}$, where \emph{a} is normalization factor related to quintessence. The spherically symmetric black hole solution with quintessence was first obtained by Kiselev \citep{Kiselev:2002dx}. Since then there were several attempts to investigate the thermodynamics of the black holes with quintessence \citep{WeiYi2011,Li:2014ixn,QuienRNThomas2012,Tharanath:2013jt,Saleh:2017vui}. We focus on the Joule-Thomson expansion of regular Bardeen black hole with a quintessence.

The paper is organized as follows. In section \ref{Action}, we explain the action of regular Bardeen black hole with quintessence. In the next section (\ref{Stress}) we discuss the stress energy tensor of regular Black hole. In section \ref{Thermodynamics} we discuss the phase transition of regular Bardeen AdS black hole. In section  \ref{JT}, we investigate Joule-Thomson expansion of the black hole. Finally, we conclude the paper in section \ref{conclusions}.




\section{R\lowercase{egular} B\lowercase{ardeen} A\lowercase{d}S \lowercase{black hole surrounded with quintessence}}
\label{Action}
The action for regular black hole in AdS spacetime coupled with nonlinear electrodynamics is given by \citep{AyonBeato:1998ub},
\begin{align}
S=\int d^4 x \sqrt{-g}\left(\frac{1}{16 \pi}R+\frac{1}{8\pi}\frac{3}{l^2} -\frac{1}{4 \pi}\mathcal{L}\left(\mathcal{F}\right) \right).
\end{align}
Where $R$ is the Ricci scalar, $l$ is the radius of the AdS space, and $\mathcal{L}\left(\mathcal{F}\right)$ represents the Lagrangian for a nonlinear electrodynamics source,
\begin{align}
\mathcal{L}\left(\mathcal{F} \right)=\frac{3M}{\beta^3}\left(\frac{ \sqrt{4\beta ^2 \mathcal{F}}}{1+\sqrt{4\beta ^2 \mathcal{F}}}\right).
\end{align}
where $\mathcal{F}=F_{\mu\nu}F^{\mu\nu}$, ${F}_{\mu\nu}$ is the strength of the electromagnetic field,  $\beta$ is the magnetic monopole charge . The solution for this action is given by \citep{AyonBeato:1998ub}
\begin{align*}
\label{metric1}
ds^2&= -\left(1-\frac{2 \mathcal{M}(r)}{r} +\frac{r^2}{l^2}\right)dt^2+ \frac{dr^2}{\left(1-\frac{2 \mathcal{M}(r)}{r}+\frac{r^2}{l^2} \right)} +r^2d\theta^2 +r^2 \sin^2 \theta d\phi^2,
\end{align*}
with $\mathcal{M}(r)= \frac{Mr^3}{(r^2+\beta ^2)^{3/2}}$.
The presence of quintessence throughout the universe makes it important to probe its effects on the black holes. Kislev's phenomenological model is used to construct a regular-Bardeen black hole surrounded by quintessence. One can obtain the metric with the effect of quintessence on regular Bardeen black hole by solving Einstein equations. It is given by \citep{kiselev2003quintessence,li2014effects,fan2017critical,saleh2018thermodynamics}
 \begin{align*}
ds^2&= -f(r)dt^2+ \frac{dr^2}{f(r)} +r^2d\theta^2 +r^2 \sin^2 \theta d\phi^2,
\end{align*}
where
\begin{eqnarray}
 \label{metric2}
 f\left(r\right)= \left(1-\frac{2 M r^2}{ \left(\beta^2+r^2\right)^{3/2}} +\frac{r^2}{l^2}-\frac{a}{r^{3\omega_q +1}}\right),
 \end{eqnarray}
 and $\omega_q$ is the state parameter.    
\section{S\lowercase{tress-energy tensor for regular} B\lowercase{ardeen quintessence black hole}}\label{Stress}
Next, we calculate the stress-energy tensor for the regular Bardeen quintessence black holes using the formalism of tetrads. The tetrads for the metric (\ref{metric2}) are expressed as $e^{(a)}_\mu=\text{diag}(\sqrt{f(r)},\frac{1}{\sqrt{f(r)}},r,r\sin\theta)$. Using an  orthonormal frame,  we can easily calculate the stress-energy tensor for the spacetime described by equation (\ref{metric2})  as
\begin{eqnarray}
G_{\hat{t}\hat{t}}&=&-\frac{3}{8\pi}\left(\frac{a\omega_q}{r^{3\left(\omega_q+1\right)}}-\frac{2M\beta^2}{\left(r^2+\beta^2\right)^{5/2}}\right)=-G_{\hat{r}\hat{r}},\nonumber\\
G_{\hat{\theta}\hat{\theta}}&=&-\frac{3}{16\pi}\left(\frac{a\omega_q\left(1+3\omega_q\right)}{r^{3\left(\omega_q+1\right)}}-\frac{2M\beta^2\left(3r^2-2\beta^2\right)}{\left(r^2+\beta^2\right)^{7/2}}\right)=G_{\hat{\phi}\hat{\phi}}.
\end{eqnarray}
Hence, we can write 
\begin{eqnarray}
\label{prho}
\rho&=&-p_r=-\frac{3}{8\pi}\left(\frac{a\omega_q}{r^{3\left(\omega_q+1\right)}}-\frac{2M\beta^2}{\left(r^2+\beta^2\right)^{5/2}}\right),\nonumber\\
p_t&=&-\frac{3}{16\pi}\left(\frac{a\omega_q\left(1+3\omega_q\right)}{r^{3\left(\omega_q+1\right)}}-\frac{2M\beta^2\left(3r^2-2\beta^2\right)}{\left(r^2+\beta^2\right)^{7/2}}\right),
\end{eqnarray}
where $\rho$ and $p_r$ are, respectively, the energy density and the pressure in the radial direction while $p_t$ is the tangential pressure of the total energy-momentum tensor of the regular Bardeen black hole surrounded by a static quintessence field. The Eqs.~(\ref{prho}) show that the required anisotropy is restored in the expressions for $\rho$, $p_r$ and $p_t$ as $p_r\neq p_t$. This is not isotropic, so it is not a perfect fluid. For the average pressure we have
\begin{eqnarray}
\label{avg_p}
\bar{p}=-\frac{1}{8\pi}\left(\frac{3a\omega_q^2}{r^{3\left(\omega_q+1\right)}}-\frac{2M\beta^2\left(2r^2-3\beta^2\right)}{\left(r^2+\beta^2\right)^{7/2}}\right),
\end{eqnarray}
along with this we can get the effective equation of state
\begin{eqnarray}
\label{eos1}
\omega_{eff}=\frac{-\frac{3a\omega_q^2}{r^{3\left(\omega_q+1\right)}}+\frac{2M\beta^2\left(2r^2-3\beta^2\right)}{\left(r^2+\beta^2\right)^{7/2}}}{-\frac{3a\omega_q}{r^{3\left(\omega_q+1\right)}}+\frac{6M\beta^2}{\left(r^2+\beta^2\right)^{5/2}}}.
\end{eqnarray}
For the pressure ratio and relative pressure anisotropy we explicitly have
\begin{eqnarray}
\label{prr_anis}
\frac{p_t}{p_r}&=&\frac{-\frac{a\omega_q\left(1+3\omega_q\right)}{r^{3\left(\omega_q+1\right)}}+\frac{2M\beta^2\left(3r^2-2\beta^2\right)}{\left(r^2+\beta^2\right)^{7/2}}}{\frac{2a\omega_q}{r^{3\left(\omega_q+1\right)}}-\frac{4M\beta^2}{\left(r^2+\beta^2\right)^{5/2}}},\nonumber\\
\delta{p}&=&\frac{\frac{9a\omega_q\left(1+\omega_q\right)}{r^{3\left(\omega_q+1\right)}}-\frac{15M\beta^2 r^2}{\left(r^2+\beta^2\right)^{7/2}}}{-\frac{3a\omega_q^2}{r^{3\left(\omega_q+1\right)}}+\frac{2M\beta^2\left(2r^2-3\beta^2\right)}{\left(r^2+\beta^2\right)^{7/2}}}.
\end{eqnarray}
The expressions for $\omega_{eff}$, $p_t/p_r$, and $\delta{p}$ all have dependence on the radial coordinate, hence does not correspond to the perfect fluid. However, for $\beta=0$, our results go over to the expressions for the anisotropic, and imperfect fluid quintessence black holes \citep{Visser:2019brz}. 
\section{T\lowercase{hermodynamics}}\label{Thermodynamics}
One can obtain the mass of the black hole from the condition $f(r_+)=0$ at the event horizon $r_+$ as
\begin{equation}
M=\frac{1}{6} \left(\beta^2+r_+^2\right)^{3/2} \left(\frac{3 \left(l^2+r_+^2\right) }{ r_+^2 l^2}-\frac{3 a}{ r_+^{3 (w_q+1)}}\right).
\end{equation}
When $\Lambda$ is considered as thermodynamic pressure in the extended phase space (and the conjugate quantity as volume) we have
\begin{equation}
P=-\frac{\Lambda}{8\pi}=\frac{3}{8 \pi l^2}.
\end{equation} 
The presence of quintessence modifies the first law as
\begin{eqnarray}
dM=TdS+\Psi d\beta+VdP+\mathcal{A}d a .\label{eqn:First law}
\end{eqnarray}
Where $\Psi$ and $\mathcal{A}$  are the conjugate quantities corresopnds to  the non-linear magnetic  charge $\beta$ and quintessence parameter $a$ respectively.
\begin{equation}
\mathcal{A}=\left( \frac{\partial M}{\partial a}\right) _{S,\beta , P}=-\frac{1}{2r_h^{3\omega_q}}.
\end{equation}
The  quintessence parameters $\mathcal{A}$ and $a$ are playing the same role as pressure and volume in the first law to make it consistent with the Smarr relation. From first law of thermodynamics we calculate volume.
\begin{equation}
V=\left(\frac{\partial M}{\partial P}\right)_{S,\beta}=\frac{4}{3} \pi  \left(\beta^2+r_+^2\right)^{3/2},
\end{equation}
which clearly shows that it is not merely the geometric volume $\frac{4}{3} \pi r_+^3$.

The temperature of the black hole is given by
\begin{equation}
T=\frac{\kappa}{2 \pi}
\end{equation}
where $\kappa$ is the surface gravity, which can be calculated from the first derivative of $f(r)$ at the horizon, $\kappa =f'(r_+)/2$. In our case we have,
\begin{equation}
T=\frac{r_+^{-3 w_q-2} \left(3 a \left(\beta^2 (w_q+1)+r_+^2 w_q\right)+r_+^{3 w_q+1} \left(-2 \beta^2+8 \pi  P r_+^4+r_+^2\right)\right)}{4 \pi  \left(\beta^2+r_+^2\right)}.\label{Temp}
\end{equation}
From this one can obtain the equation of motion
\begin{equation}
P=\frac{ \left(r_+^{3 w_q+1}\left(4 \pi  r_+ T(\beta^2+r_+^2)+2\beta^2-r_+^2\right)-3 a \left(\beta^2 (w_q+1)+r_+^2 w_q\right)\right)}{8 \pi r_+^{3 w_q+5} }.\label{Pressure}
\end{equation}
The critical values are obtained from the conditions,
\begin{equation}
\left(\frac{\partial P}{\partial r_+}\right)_T=\left(\frac{\partial ^2 P}{\partial r_+^2}\right)_T=0.
\end{equation}
The critical values for $\omega_q=-1$ are,
\begin{equation}
\textstyle{r_c=\sqrt{\frac{1}{2} \left(\sqrt{273}+15\right)} \beta,\quad T_c=-\frac{\left(\sqrt{273}-17\right) \sqrt{\frac{1}{2} \left(\sqrt{273}+15\right)}}{24 \pi  \beta},\quad
 P_c=\frac{27 \left(5 \sqrt{273}+83\right) a \beta^2+\sqrt{273}+27}{12 \left(\sqrt{273}+15\right)^2 \pi \beta^2}}.
\end{equation}
And similarly for $\omega_q=-1/3$ we have,
\begin{equation}
\textstyle{r_c=\sqrt{\frac{1}{2} \left(\sqrt{273}+15\right)} \beta,\quad T_c=\frac{\left(\sqrt{273}-17\right) \sqrt{\frac{1}{2} \left(\sqrt{273}+15\right)} (a-1)}{24 \pi  \beta},\quad P_c=-\frac{\left(\sqrt{273}+27\right) (a-1)}{12 \left(\sqrt{273}+15\right)^2 \pi  \beta^2}}.
\end{equation}
It is clear that the critical parameters depends on the quintessence parameters.
\section{J\lowercase{oule} T\lowercase{homson} \lowercase{expansion}}
\label{JT}
Joule Thomson expansion in thermodynamics is an irreversible  process, which explains the temperature change of a gas when it is passing through a porous plug from high pressure region to low pressure region. Enthalpy of the system remains constant during this expansion. The set of values $(T,P)$ throughout the process with the constraint of constant enthalpy defines the isenthalpic curve. The expression for Joule Thomson coefficient (slope of the isenthalpic curve) is given by,
\begin{equation}
\mu_j=\left(\frac{\partial T}{\partial P}\right)_H.
\end{equation}
The sign of this coefficient tells about the cooling and heating phases. During the expansion pressure always decreases. The coefficient of Joule Thomson expansion can also be written as,
\begin{equation}
\mu_j=\left(\frac{\partial T}{\partial P}\right)_M=\frac{1}{C_P}\left[T\left(\frac{\partial V}{\partial T}\right)_P-V\right].
\end{equation}
From this one can get the inversion temperature by setting $\mu_j=0$,
\begin{equation}
T_i=V\left(\frac{\partial T}{\partial V}\right)_P.
\end{equation}
This is at the maxima of the isenthaplic curve, with the corresponding inversion pressure. The point defined by inversion temperature and inversion pressure is called inversion point.

\begin{figure}[ht!]
    \subfigure[$a=1/4$ , $\omega_q =-1$]
     {
\includegraphics[width=0.35\textwidth]{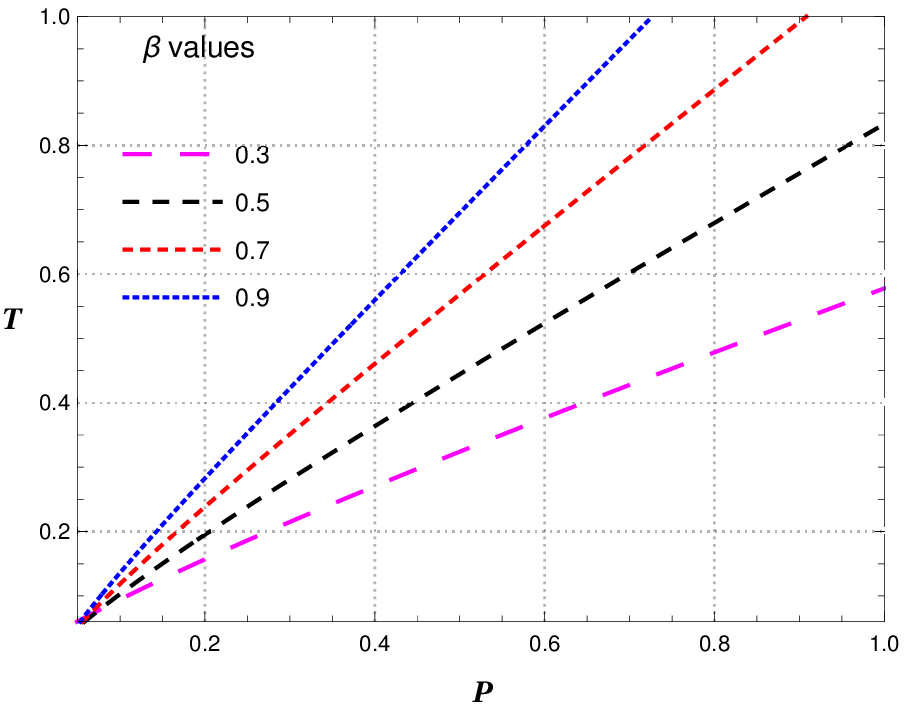}
\label{fig:a=1by4,w=-1}
    }
    \subfigure[$a=1/4$ , $\omega_q =-1/3$]
     {
\includegraphics[width=0.35\textwidth]{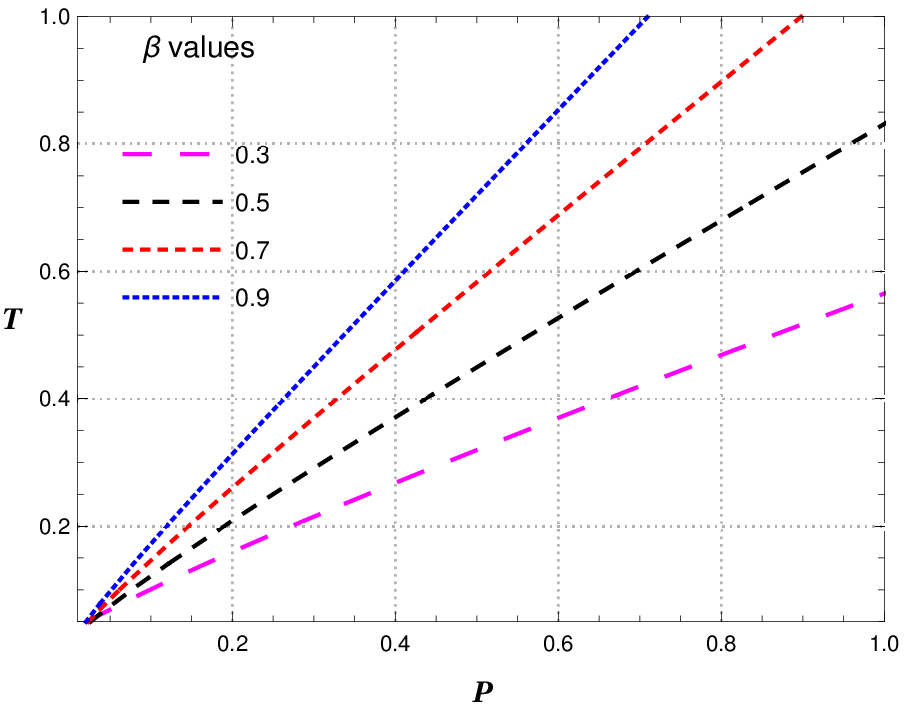}
\label{fig:a=1by4,w=-1/3}
    }    
\subfigure[$a=1$ , $\omega_q =-1$]
     {
\includegraphics[width=0.35\textwidth]{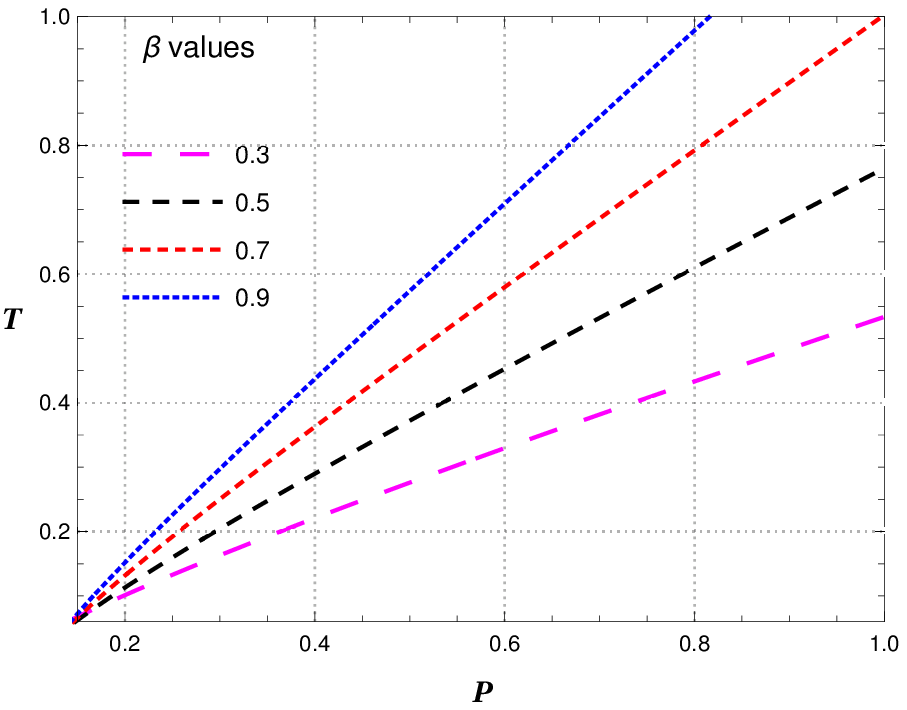}
\label{fig:a=1,w=-1}
    }
    \subfigure[$a=1$ , $\omega_q =-1/3$]
     {
\includegraphics[width=0.35\textwidth]{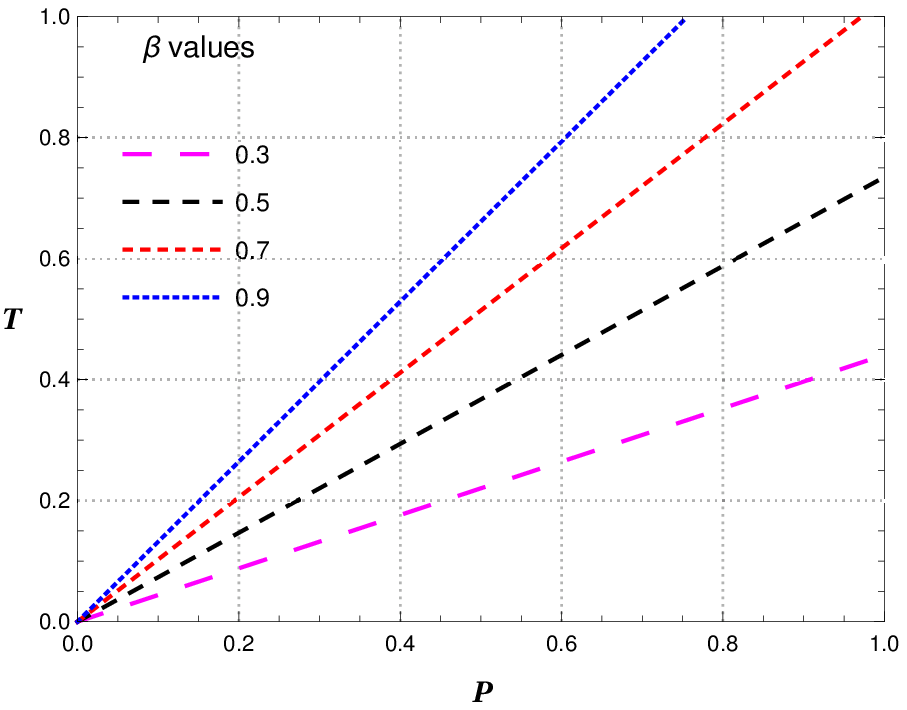}
\label{fig:a=1,w=-1/3}
    }    
     \\
 \caption{Inversion curve for different values of $\beta$.}
 \label{linear_plot}
 \end{figure}

The Joule Thomson expansion is also studied in AdS black hole \citep{okcu2017joule} with the same definitions of isenthalpic curve and inversion temperature. This is evident from the similarity between the van der Waals system and black hole system in the extended phase space. The pressure $P$  and temperature $T$ of the regular Bardeen black hole with quintessence  in terms of $M$ and $r_+$ are,
\begin{equation}
P(M,r_+)=\frac{3 \left(a r_+^{-3 w_q-1}+\frac{2 M r_+^2}{\left(\beta^2+r_+^2\right)^{3/2}}-1\right)}{8 \pi r_+^2},\label{P_in_M}
\end{equation}
and
\begin{equation}
T(M,r_+)=\frac{3 a (w_q+1) r_+^{-3 w_q}+\frac{6 M r_+^5}{\left(\beta^2+r_+^2\right)^{5/2}}-2 r_+}{4 \pi  r_+^2}.\label{T_in_M}
\end{equation}
Using the above equations (\ref{P_in_M}) and (\ref{T_in_M}), we plot the isenthalpic curves for the regular Bardeen black hole. 

\begin{figure}[ht!]
         \subfigure[$\beta=0.9$ , $a=1/4$]
          {
   \includegraphics[width=0.35\textwidth]{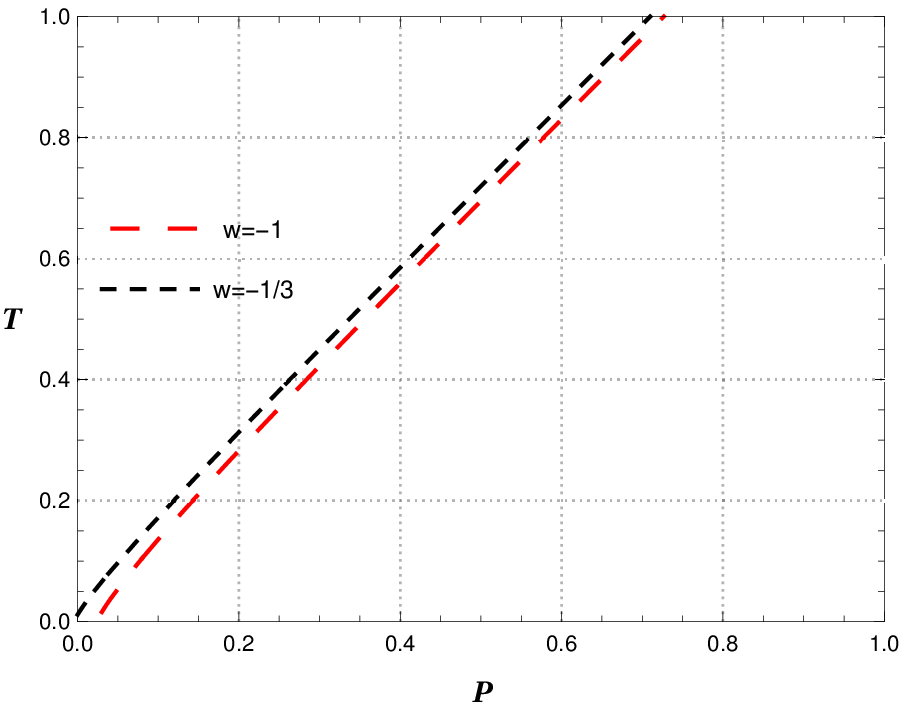}
    \label{fig:b=0.9,a=1/4}
                         }
 \subfigure[$\beta=0.9$ , $a=1$]
      {
 \includegraphics[width=0.35\textwidth]{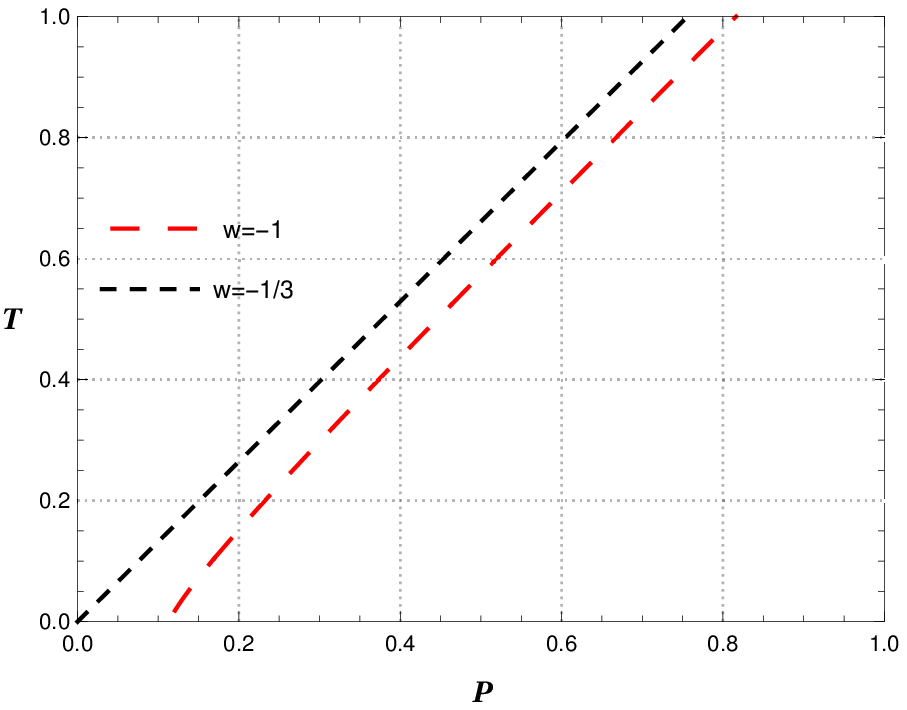}
 \label{fig:b=0.9,a=1}
     }    
               \subfigure[$\beta=0.9$ , $\omega_q=-1/3$]
                      {
  \includegraphics[width=0.35\textwidth]{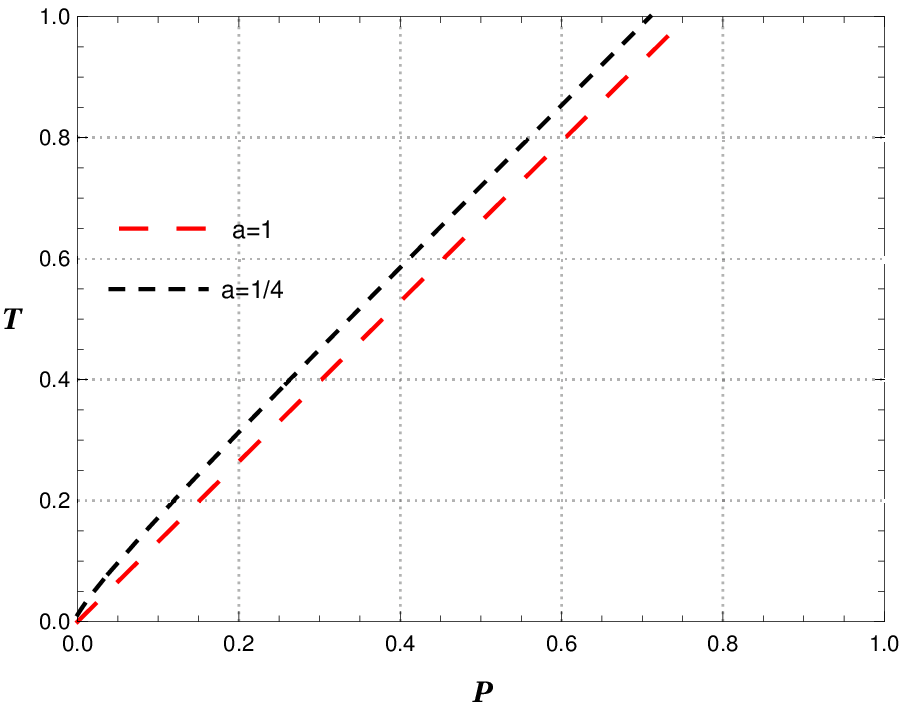}
                     \label{fig:b=0.9,w=-1/3}
                         }
 \subfigure[$\beta=0.9$ , $\omega_q=-1$]
      {
 \includegraphics[width=0.35\textwidth]{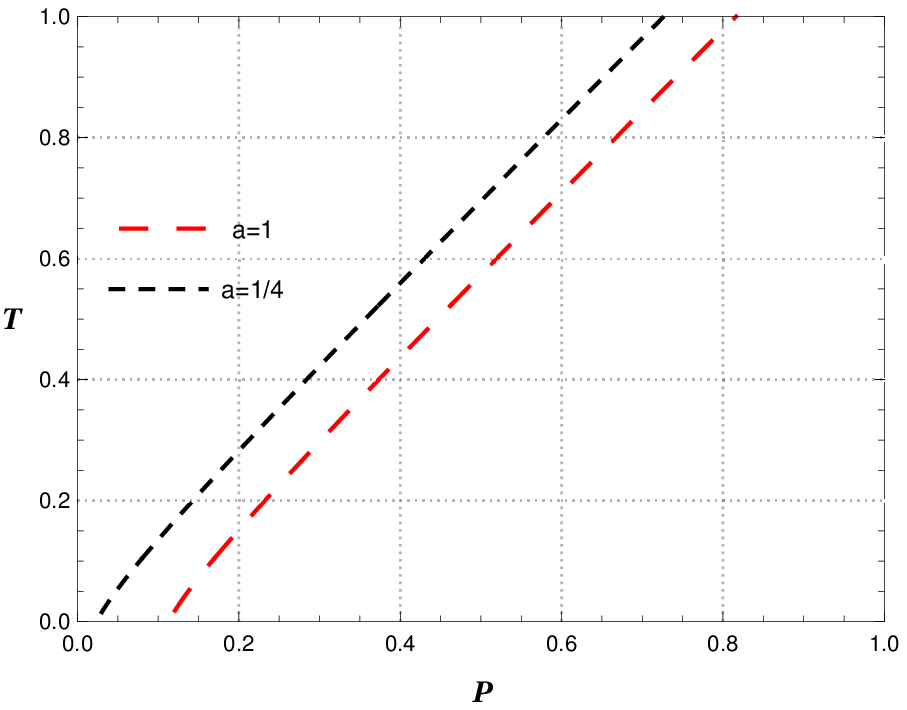}
 \label{fig:b=0.9,w=-1}
     }                         
     \\
 \caption{Inversion curve for different values of quintessence parameters $\omega _q$ and $a$.}\label{Inversion varying w}
 \end{figure}
The Joule Thomson coefficient for the black hole under consideration is,
\begin{equation}
\mu _j =\textstyle{ \frac{2 r \left(3 a \left(\beta ^4 \left(3 w_q ^2+5 w_q +2\right)+r^4 w_q  (3 w_q +5)+\beta ^2 r^2 \left(6 w_q ^2+10 w_q +7\right)\right)+r^{3 w_q +1} \left(-2 \beta ^4+16 \pi  P r^6+r^4 \left(4-24 \pi  \beta ^2 P\right)-13 \beta ^2 r^2\right)\right)}{3 \left(\beta ^2+r^2\right) \left(3 a \left(\beta ^2 (w_q +1)+r^2 w_q \right)+r^{3 w_q +1} \left(-2 \beta ^2+8 \pi  P r^4+r^2\right)\right)}}.
\label{muj}
\end{equation}
And by setting $\mu_j=0$ the inversion temperature is,
\begin{equation}
\textstyle{T_i=\frac{r_+^{-3 w_q-4} \left(r_+^{3 w_q+1} \left(2 \beta^4+\beta^2 r_+^2 \left(24 \pi  P r_+^2+7\right)+r_+^4 \left(8 \pi  P r_+^2-1\right)\right)-3 a \left(\beta^4 \left(3 w_q^2+5 w_q+2\right)+\beta^2 r_+^2 \left(6 w_q^2+7 w_q+4\right)+r_+^4 w_q (3 w_q+2)\right)\right)}{12 \pi  \left(\beta^2+r_+^2\right)}}.
\label{Ti}
\end{equation}
Setting $\mu _j=0$ in equation (\ref{muj}) we solve for $r_+$. From that choosing an appropriate root and substituting in equation (\ref{Temp}) we plot the inversion curve in the $T-P$ plane, which is shown in the figure \ref{linear_plot} and \ref{Inversion varying w}.

We have studied the inversion curves for different values of monopole charge $\beta$ in figure \ref{linear_plot}. In all the four plots the position of inversion point $(T_i,P_i)$ shifts to higher values with increase in charge $\beta$. The effect of quintessence parameters $\omega_q$ and $a$ are explicitly depicted in figure \ref{Inversion varying w}. In the figures  \ref{fig:b=0.9,a=1/4} and \ref{fig:b=0.9,a=1} we plotted inversion curves with different quintessence state parameters, shows that increase of $w_q$ from $-1$ to $-1/3$ increases the inversion temperature. They also show that the separation between the inversion curves for different values of $\omega _q$ depends on $a$. In the second set in figure \ref{Inversion varying w} (the curves  \ref{fig:b=0.9,w=-1/3} and \ref{fig:b=0.9,w=-1}), we have plotted the inversion curves by varying the quintessence normalization constant \emph{a}. These plots show a minor decrease in the inversion points with increase in $a$. In summary, the slope of the inversion curve increases sharply for the change in charge of $\beta$. But the same is not observed for the change in quintessence parameters $w_q$ and \emph{a}. The slope of the inversion curve remains same in this case.

\begin{figure}
    \subfigure[$\beta=0.3$, $a=1/4$, $\omega=-1$]
     {
\includegraphics[width=0.35\textwidth]{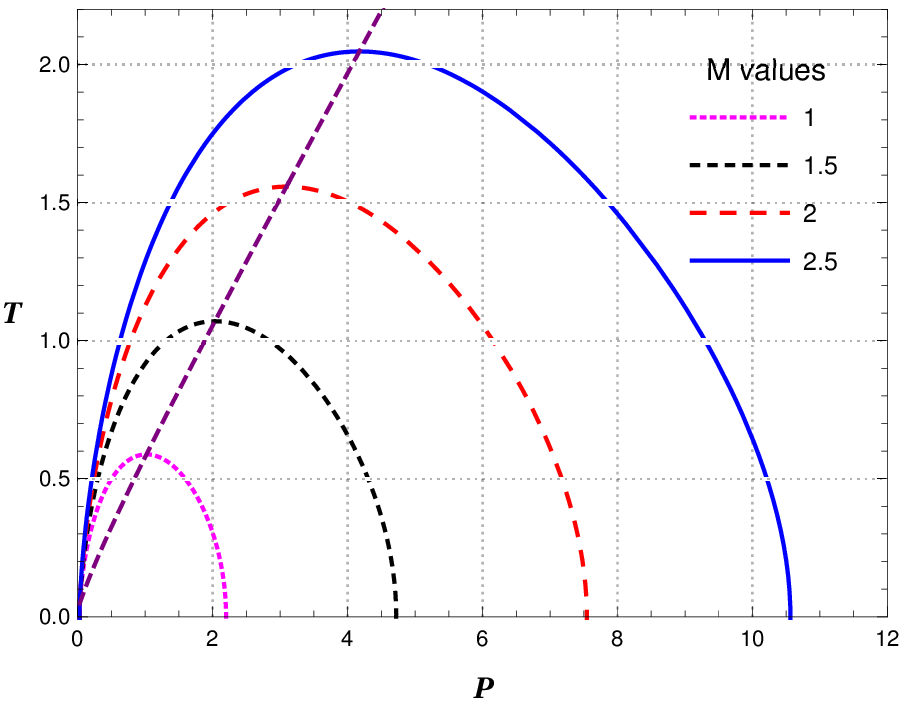}
\label{set1a}
    }
\subfigure[$\beta=0.5$, $a=1/4$, $\omega=-1$]
     {
\includegraphics[width=0.35\textwidth]{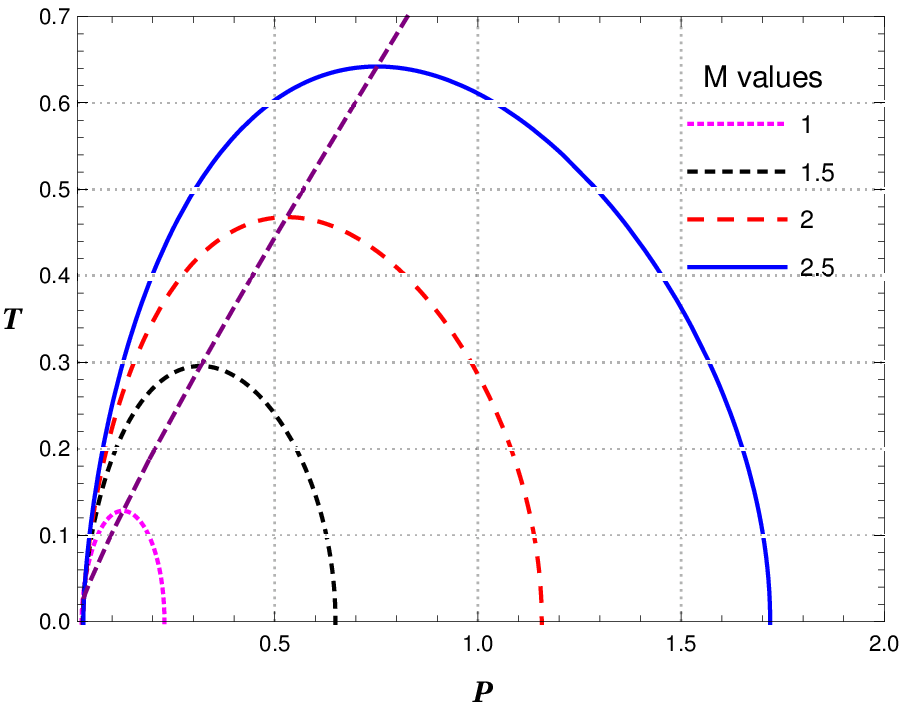}
\label{set1b}
    }
    
    \subfigure[$\beta=0.7$, $a=1/4$, $\omega=-1$]
         {
    \includegraphics[width=0.35\textwidth]{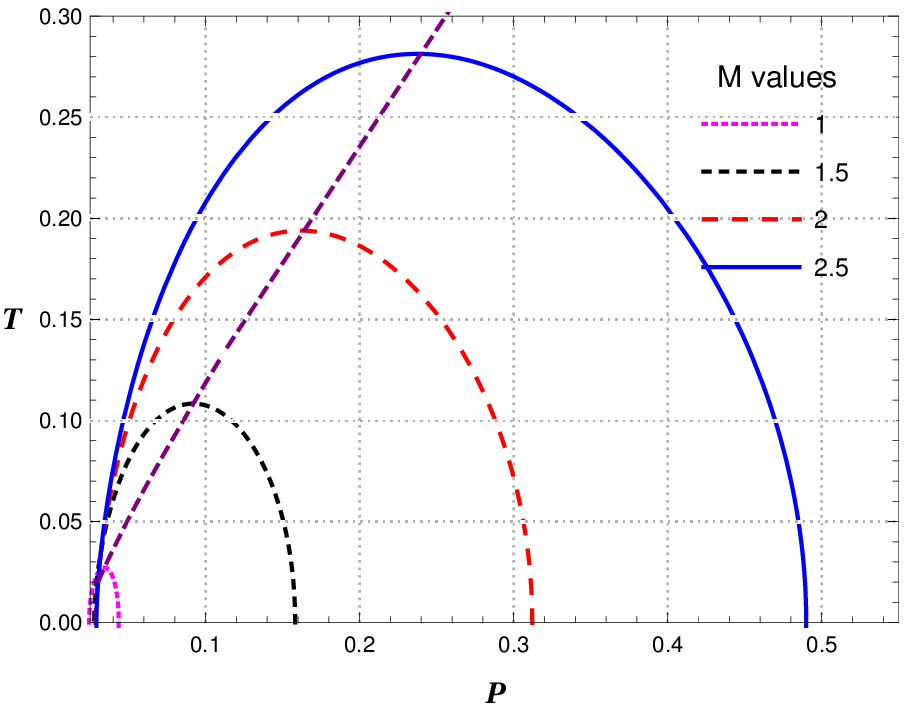}
    \label{set1c}
        }
        \subfigure[$\beta=0.9$, $a=1/4$, $\omega=-1$]
                         {  \includegraphics[width=0.35\textwidth]{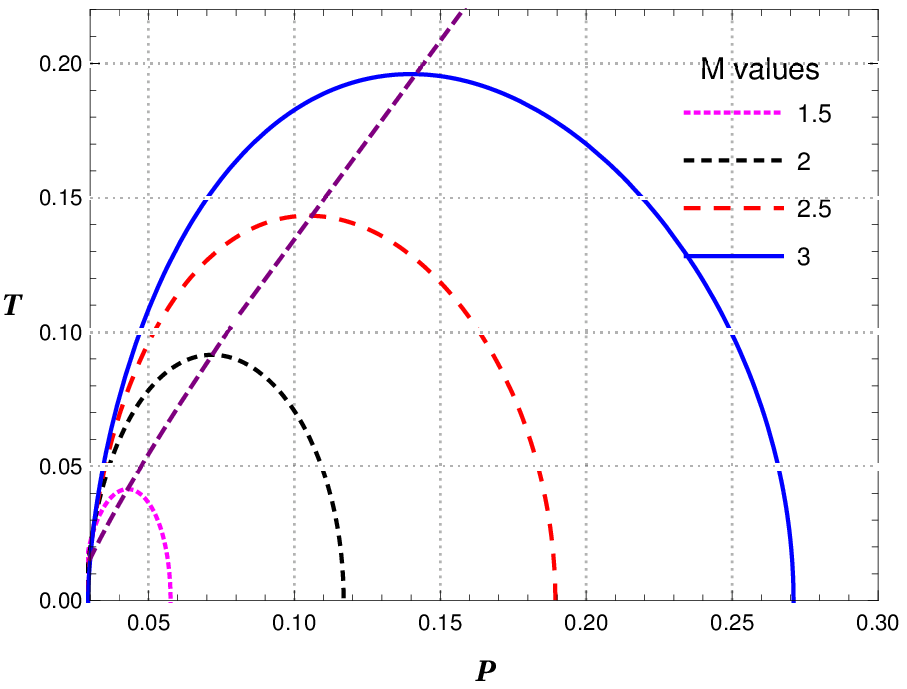}
      \label{set1d}
                        }
                          
  \subfigure[$\beta=0.3$, $a=1/4$, $\omega=-1/3$]
     {
\includegraphics[width=0.35\textwidth]{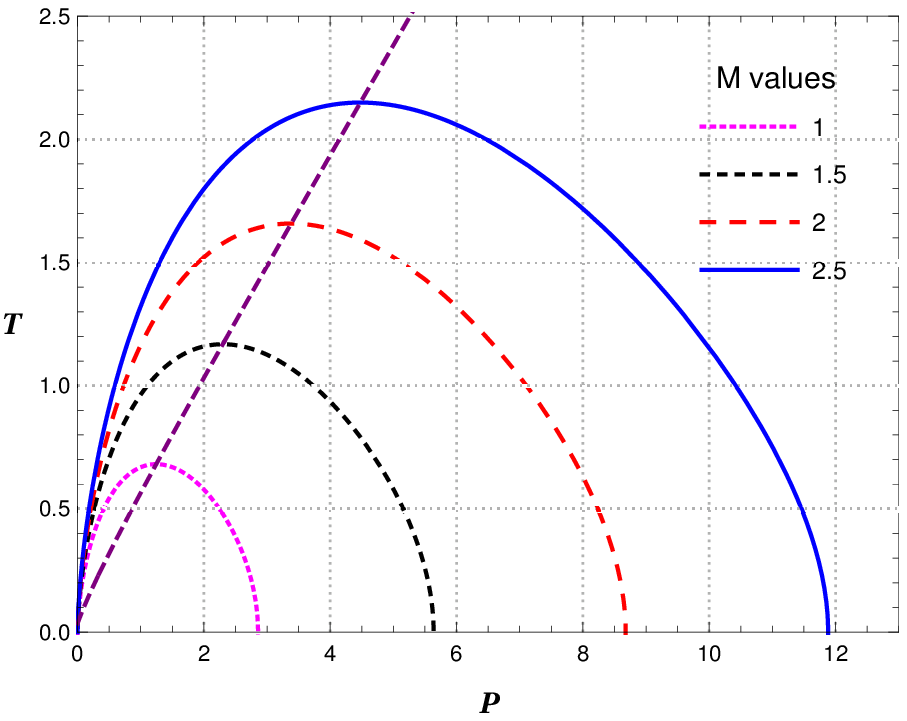}
\label{set2a}
    }
\subfigure[$\beta=0.5$, $a=1/4$, $\omega=-1/3$]
     {
\includegraphics[width=0.35\textwidth]{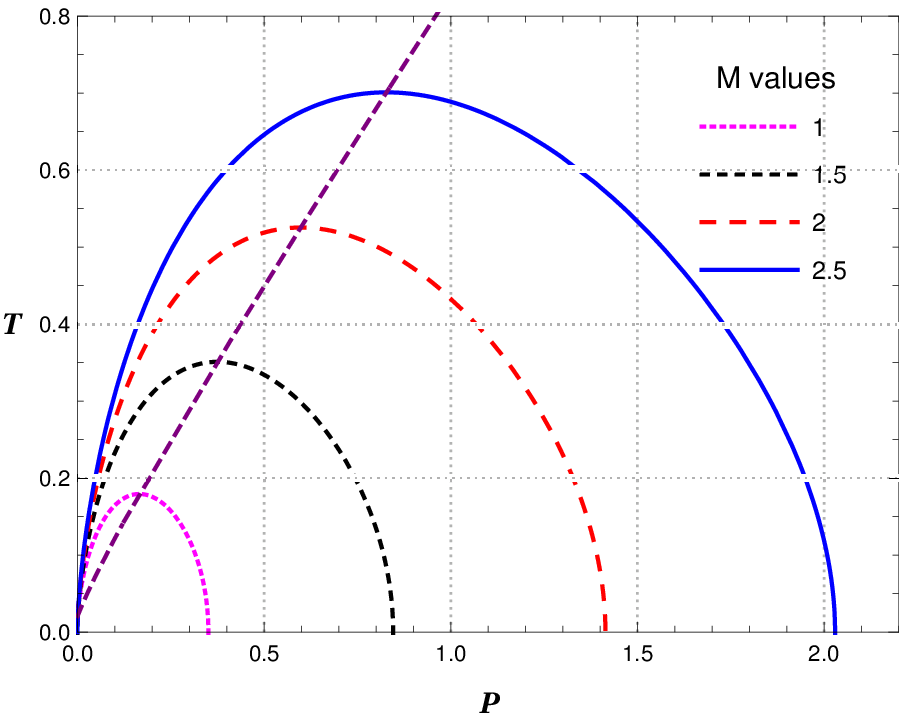}
\label{set2b}
    }
    
    \subfigure[$\beta=0.7$, $a=1/4$, $\omega=-1/3$]
         {
    \includegraphics[width=0.35\textwidth]{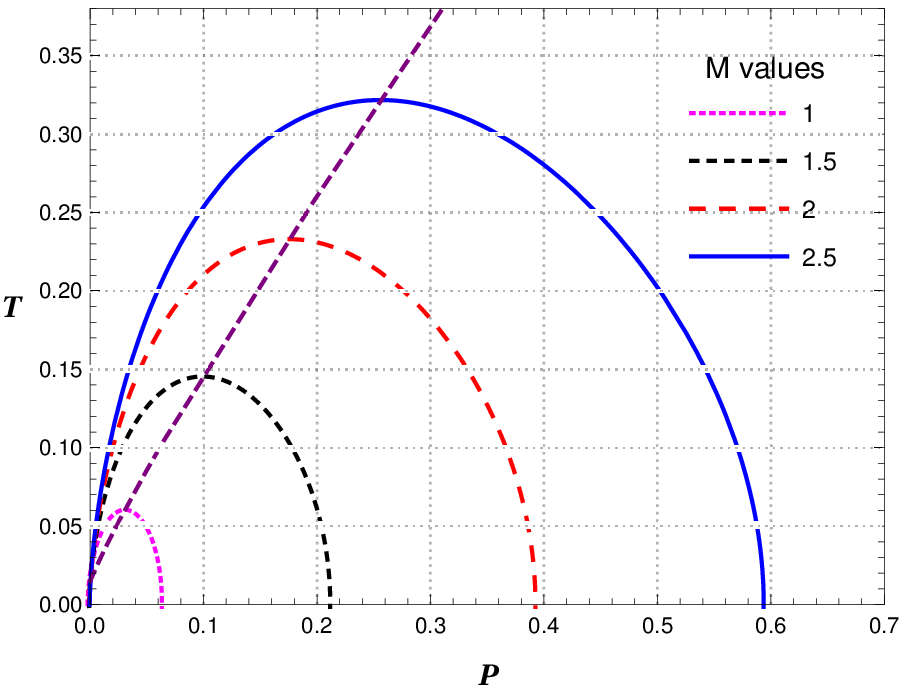}
    \label{set2c}
        }
     \subfigure[$\beta=0.9$, $a=1/4$, $\omega=-1/3$]
                         {
     \includegraphics[width=0.35\textwidth]{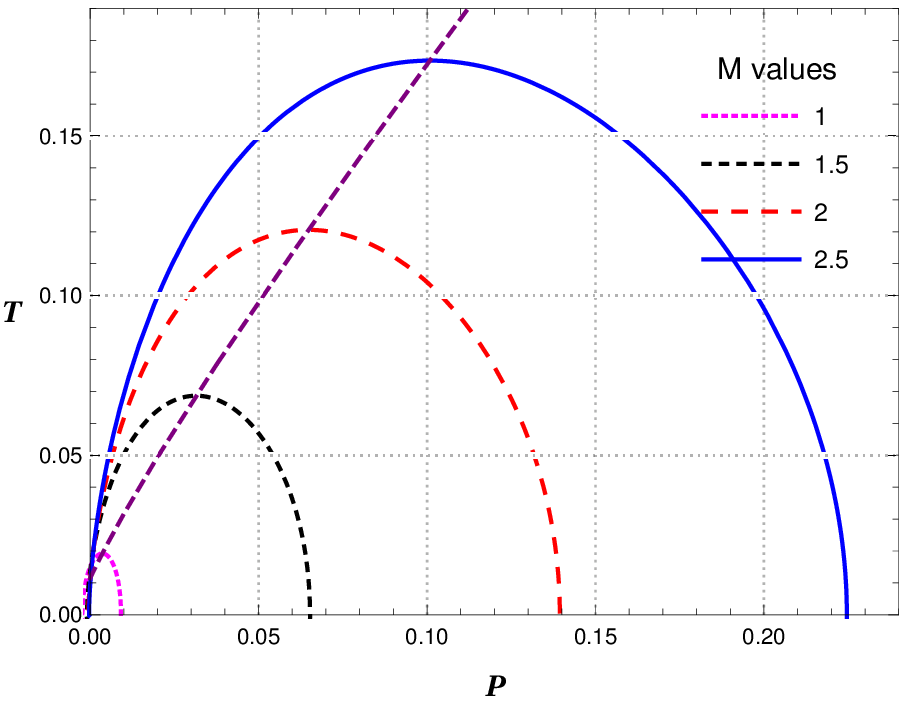}
    \label{set2d}
                        }
        \\
 \caption{Isenthalpic curves for different values of mass. The varition with respect to $\omega_q$ for a fixed $a=1/4$. }\label{Isenthalpic-1}        
  \end{figure}

\begin{figure}[]
    \subfigure[$\beta=0.3$, $a=1$, $\omega=-1$]
     {
\includegraphics[width=0.35\textwidth]{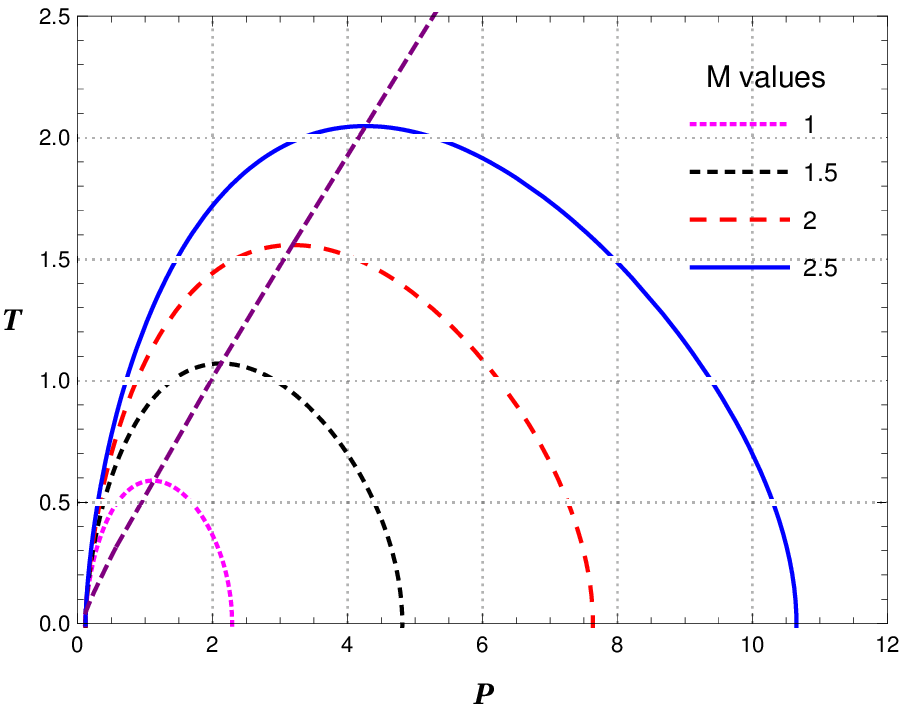}
\label{set3a}
    }
\subfigure[$\beta=0.5$, $a=1$, $\omega=-1$]
     {
\includegraphics[width=0.35\textwidth]{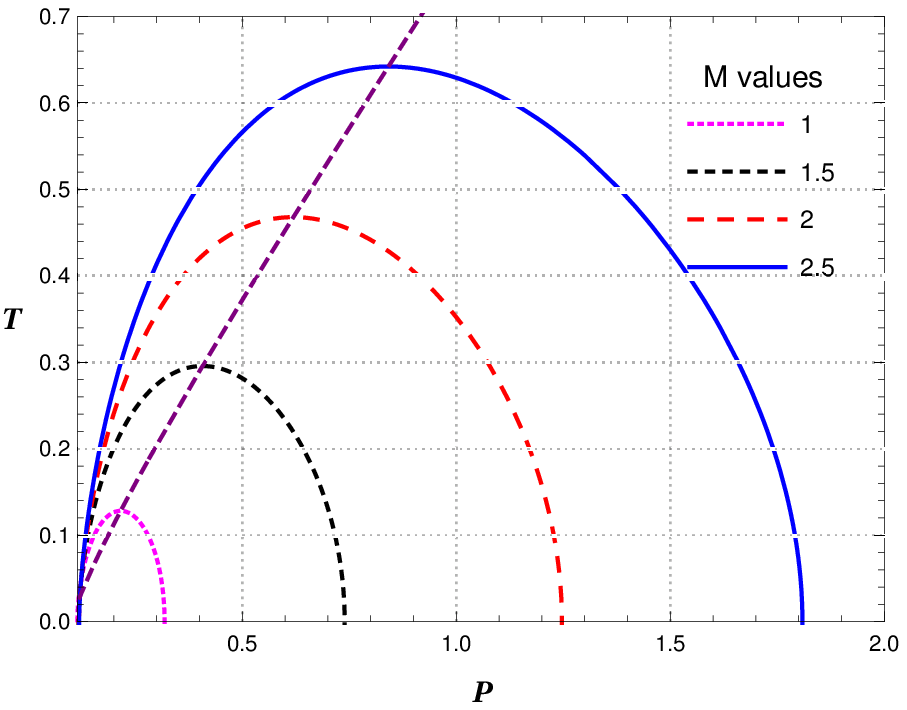}
\label{set3b}
    }
    
    \subfigure[$\beta=0.7$, $a=1$, $\omega=-1$]
         {
    \includegraphics[width=0.35\textwidth]{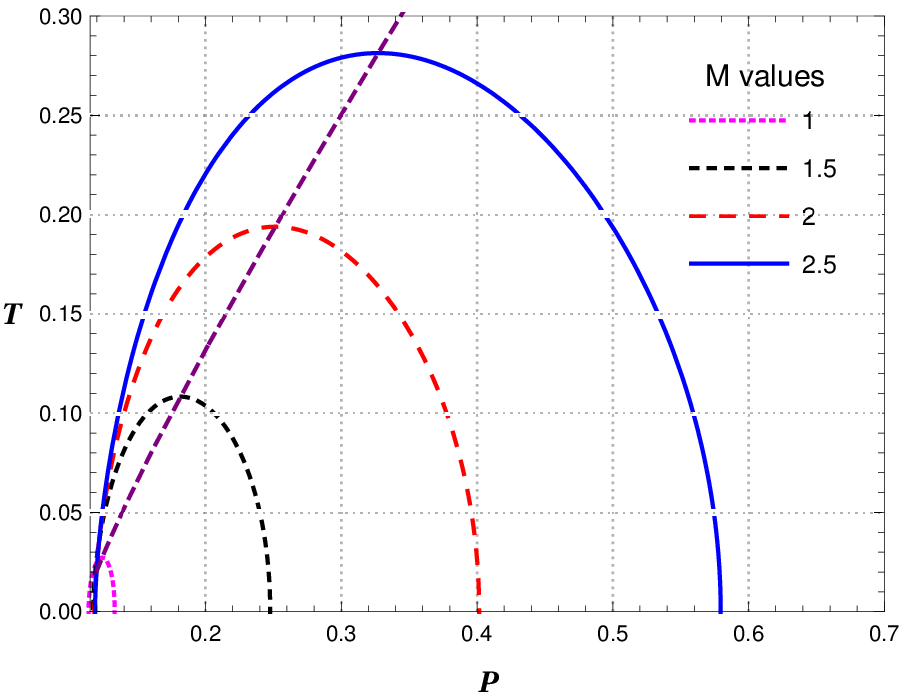}
\label{set3c}
        }
        \subfigure[$\beta=0.9$, $a=1$, $\omega=-1$]
                         {
  \includegraphics[width=0.35\textwidth]{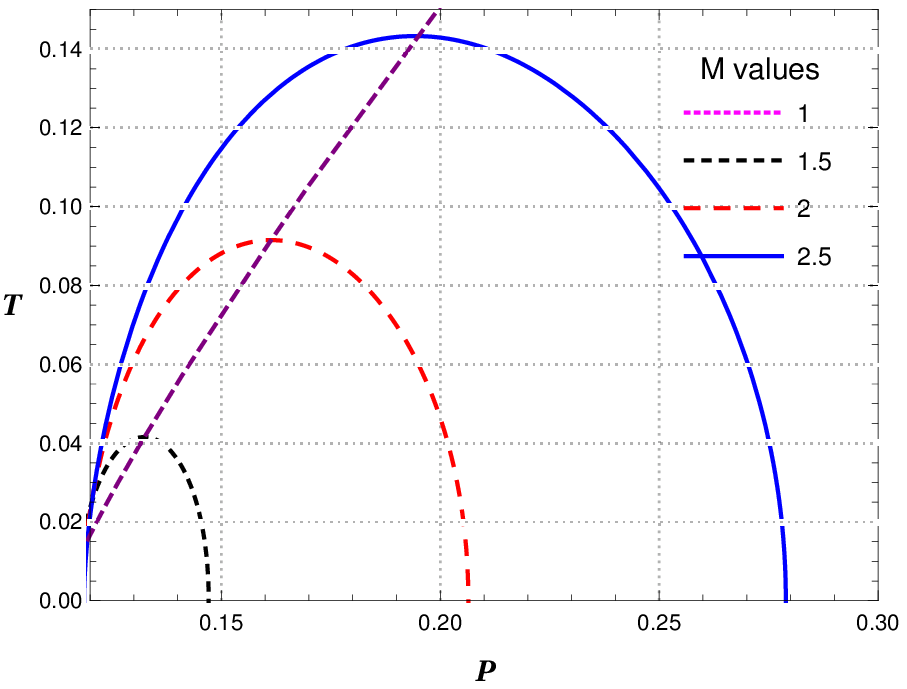}
\label{set3d}
                        }
                        
     \subfigure[$\beta=0.3$, $a=1$, $\omega=-1/3$]
      {
 \includegraphics[width=0.35\textwidth]{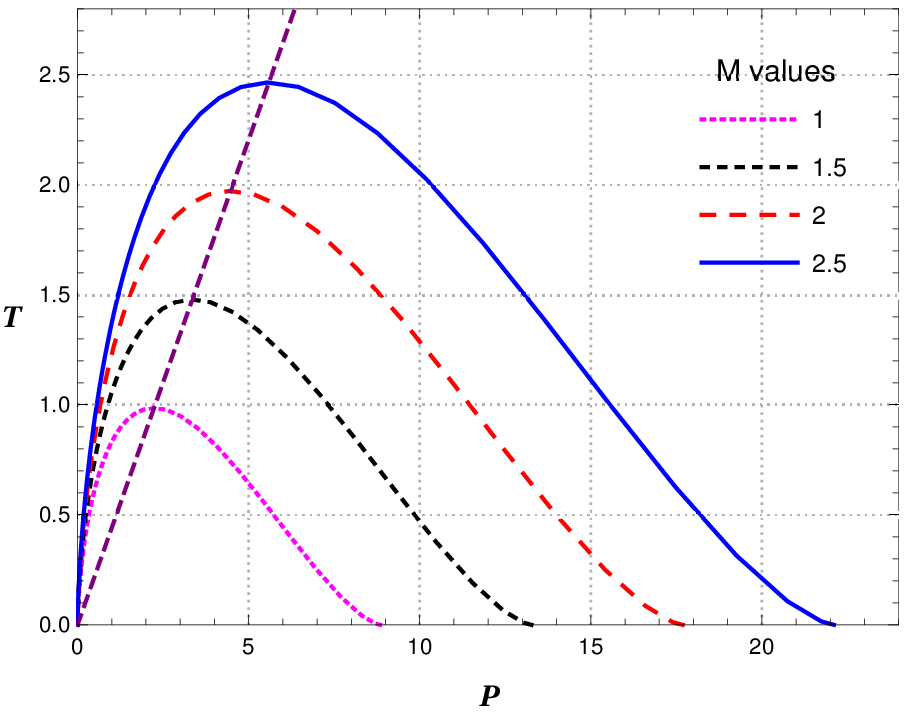}
\label{set4a}
     }
 \subfigure[$\beta=0.5$, $a=1$, $\omega=-1/3$]
      {
 \includegraphics[width=0.35\textwidth]{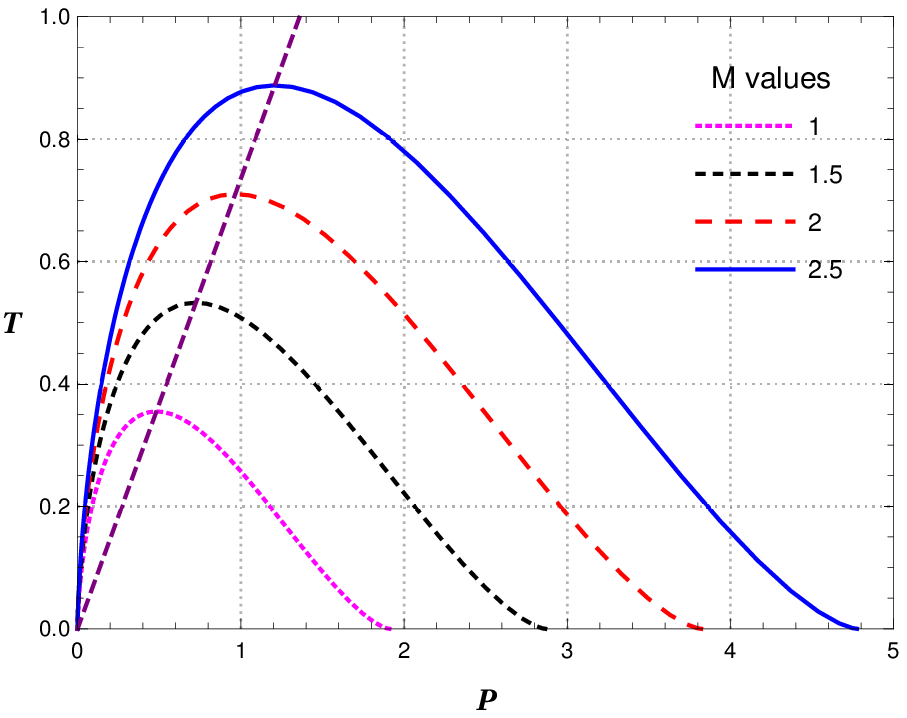}
\label{set4b}
     }
     
     \subfigure[$\beta=0.7$, $a=1$, $\omega=-1/3$]
          {
     \includegraphics[width=0.35\textwidth]{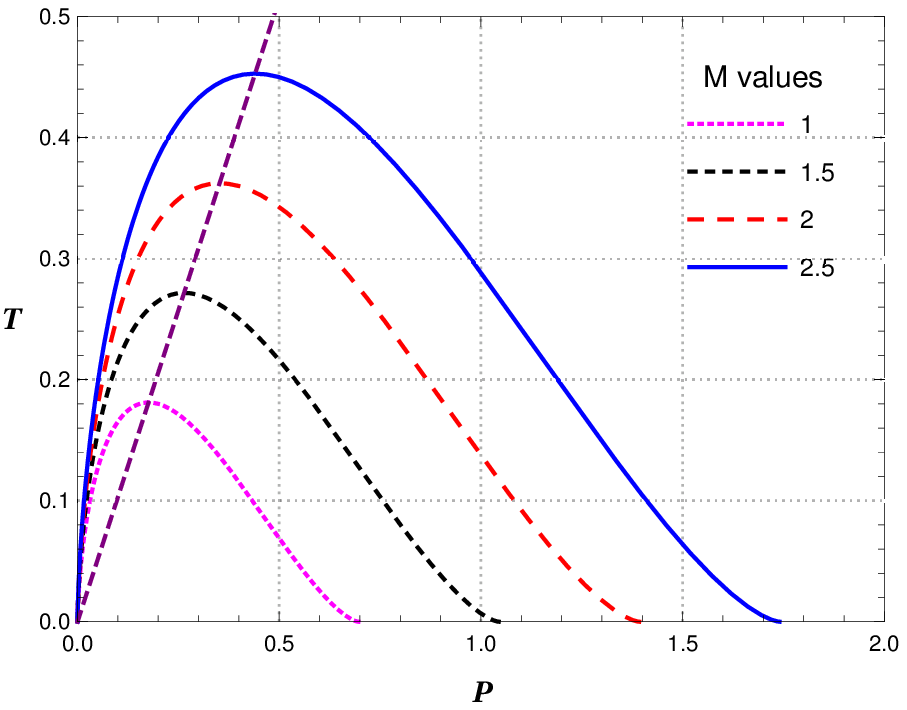}
\label{set4c}
         }
         \subfigure[$\beta=0.9$, $a=1$, $\omega=-1/3$]
                          {
     \includegraphics[width=0.35\textwidth]{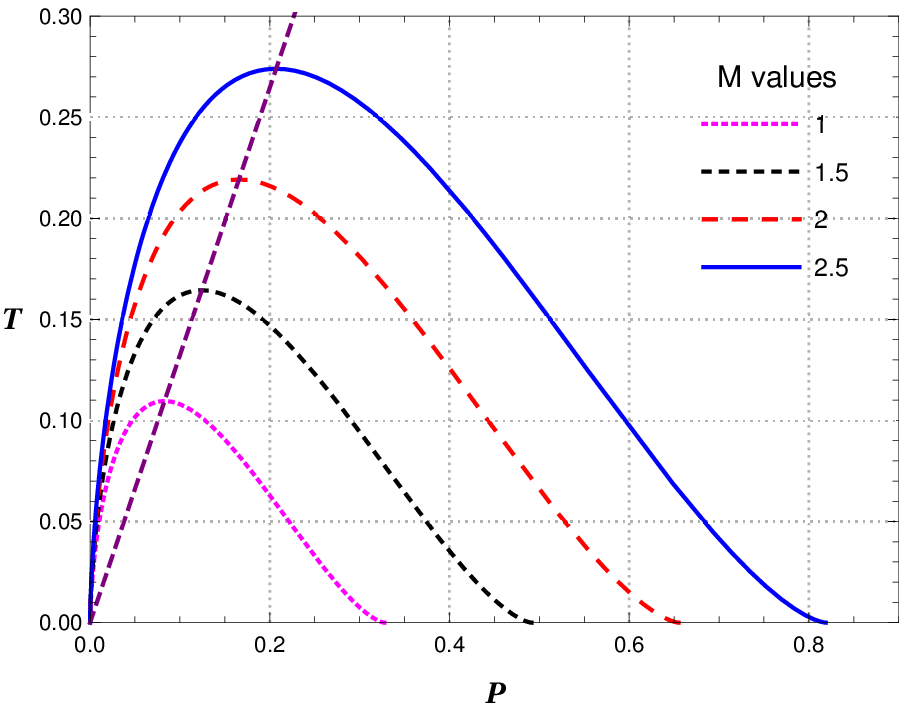}
\label{set4d}
                         }        \\
 \caption{Isenthalpic curves for different values of mass. The varition with respect to $\omega_q$ for a fixed $a=1$.  }\label{Isenthalpic-2}  
 \end{figure}

The isenthalpic (constant mass) curves are also studied for the JT expansion. We plot the isenthalpic curves on the $T-P$ plane for fixed values of enthalpy  (mass). The intersection point of the inversion curve and isenthalpic curve ($\mu=0$) separates the cooling and heating phases. And that happens at maxima of isenthalps. The figure \ref{Isenthalpic-1} and \ref{Isenthalpic-2} shows that the inversion point and isenthalpic curve depend significantly on the mass (enthalpy) of the black hole. With the increase of mass, the inversion temperature $T_i$ and pressure $P_i$ increases.

We show the effect of quintessence parameters ($\omega _q$ and $a$) and charge $\beta$ in four set of figures (set 1: figure \ref{set1a}-\ref{set1d}, set 2: figure \ref{set2a}-\ref{set2d}, set 3: figure \ref{set3a}-\ref{set3d}, set 4: figure \ref{set4a}-\ref{set4d}). In all the four sets we observe that the height of the isenthalpic curve and the value of inversion temperature reduces with increase in charge $\beta$. Comparing set 1 with 2 and 3 with 4 we see that, for an increase in quintessence state parameter $w_q$ from $-1$ to $-1/3$ there is a slight increase in the constant enthalpy curves and in the inversion temperature. The effect of quintessence normalization factor $a$ can be seen by comparing figure \ref{Isenthalpic-1} with \ref{Isenthalpic-2}. An increase in the value \emph{a} also increases the height of isenthalpic curves and the inversion temperature. A notable point here is that the isenthalpic curve for $a=1$, $\omega_q=-1/3$ case shows a deviation in shape from a usual semicircle to  a more concave form. This results in the shift of inversion pressure to lower values. These results are consistent with the earlier findings that, the thermodynamics of the black hole is affected by the quintessence field. In fact the JT expansion is an inherent feature of van der Waals fluid. And therefore the changes in thermodynamic properties and JT expansion are correlated.

\newpage
\section{Conclusions and Discussions}\label{conclusions}
In this paper, we have studied the Joule-Thomson expansion of regular Bardeen AdS black hole surrounded by static anisotropic quintessence field. The key feature that leads to  JT expansion in the black hole is, the identification of cosmological constant as pressure which enables us to redefine the black hole mass as the enthalpy. We have calculated an exact expression for the JT coefficient ($\mu$) which depends on the quintessence parameters $\omega _q$ and $a$. Furthermore, we investigated the JT expansion intuitively by using the inversion and isenthalpic curves in the $T-P$ plane. The inversion curve divides the isenthalpic curves in the $T-P$ plane into two regions. The upper region leads to cooling and the lower region results in heating, in the final state of the JT expansion. Nevertheless, the slope of the inversion curve always remains positive. 
 
We have studied the isenthalpic curves and the inversion curves for different values of enthalpy (mass), charge $\beta$ and quintessence parameters (\emph{a}, $w_q$), separately. The constant enthalpy curves show that the inversion temperature $T_i$ increases for the larger enthalpy and reduces with increase in the charge $\beta$. It is observed that the slope of inversion curves increases with the charge $\beta$. However the quintessence dark energy affects the inversion and isenthalpic curves significantly, which is an interesting result from our study. Both the quintessence parameters influences the JT expansion in same manner. Increase in the value of both  $\omega _q$ and $a$ increases the height of isenthalpic curves and the inversion temperature. This is the effect of static anisotropic quintessence field on the JT expansion.

This result is interesting because, the thermodynamics of black holes with quintessence depends on their quintessence parameters. Dependence of JT expansion on these parameters is intriguing as the quintessence plays the role of dark energy in many cosmological models.

\section*{Aknowledgement}
The authors R.K.V., A.R.C.L. and N.K.A. would like to thank the department of physics, National Institute of Technology Karnataka. The author N.K.A. also thank UGC, Govt. of India for financial support through SRF scheme.

 \bibliographystyle{elsarticle-num} 
  \bibliography{JT_Quintessence}
  
\end{document}